# SIMULATIONS OF HIGHER ORDER MODES IN THE ACC39 MODULE OF FLASH

I.R.R. Shinton[†\*], R.M. Jones[†‡], Z. Li[°], P. Zhang[†‡\*]

[†]School of Physics and Astronomy, The University of Manchester, Manchester, U.K.
[\*]The Cockcroft Institute of Accelerator Science and Technology, Daresbury, U.K.
[°]SLAC, Menlo Park, California, USA.
[‡]DESY, Hamburg, Germany.

*Abstract*

This study is focused on the development of a HOM-based BPM system for the ACC39 module currently installed and in operation at FLASH. A similar system is anticipated to be installed at XFEL. Coupled inter-cavity modes are simulated together with a limited band of trapped modes. A suite of finite element computer codes (including HFSS and ACE3P) and globalised scattering matrix calculations (GSM) are used to investigate the modes in these cavities with a view to providing guidance on their use as a cavity beam diagnostic.

## INTRODUCTION

At the FLASH facility in DESY a higher order mode (HOM) BPM system has been installed and used to both align the beam and provide useful cavity diagnostics since 2006 with beam based alignment to within 5μm [1]. This was based on HOMs excited in the 1.3GHz accelerating cavities

However, the HOMs in the third harmonic cavities of ACC39 propagate from one cavity to the next. Previous experimental studies of both beam and non-beam based measurements [2] have highlighted the difficulty of finding isolated modes suitable for use in a HOM BPM system. This complicates both the experimental analysis and the design of HOM BPM diagnostics. We identified the first beampipe and 5[th] dipole regions as having localised modes that may be suitable for a HOM BPM system [3]. However we were restricted to using the semi-analytical GSM technique [4] to simulate ACC39 en-masse. In this paper we explore ACC39 in its entirety using the S3P driven module of the electromagnetic parallel computational suite ACE3P that has been developed by the ACD group at SLAC [5] over the past 20 years.

## ACC39 COMPARISON OF GSM TO S3P

The naming convention of the cavities and ports of ACC39 used in this paper is depicted in Fig. 1. We will define the naming convention as follows *c1p1,c4p2* refers to the transmission from HOM p1 of cavity C1 to the HOM p2 of cavity C4.

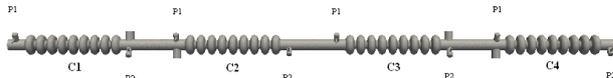

Figure 1: Parameters used in GSM structural calculation.

Here we compare the transmission ($S_{21}$) results obtained using GSM with 20 modes per port, to a similar calculation made with the S3P driven modal module of ACE3P, with a few exceptions. Firstly the bellows were not considered in the calculation (as they are computationally expensive to simulate). Secondly only a single mode, the TEM mode, was excited at each port in the S3P calculation. The mesh generated using CUBIT for use with S3P was a third order mesh consisting of more than 3 million elements.

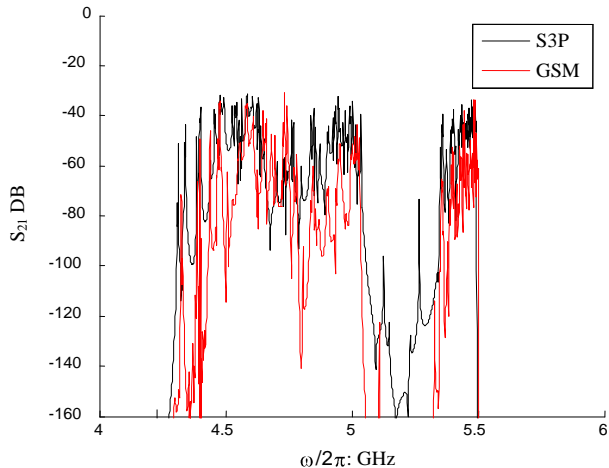

Figure 2: Comparison of $S_{21}$ calculated by GSM and S3P (ACE3P) across the entire module (*c1p1,c4p2*). In this calculation all other ports are bounded by magnetic boundary conditions, so that a direct comparison can be made with the GSM calculation.

It is evident that GSM gives comparable results to S3P simulations. However below 4.39GHz the discrepancy is larger. We attribute the discrepancy in the results as being due to modes below cut-off (the beam-pipe cut-off is ~4.39GHz for dipole modes) in the cascaded blocks. The cascading method relies on there being at least one propagating mode present in order to be accurate. This is further emphasised by GSM calculations conducted with less modes, where this disagreement in the region of the cut-off becomes more noticeable.

## ANALYSIS OF ACC39 USING S3P

To provide an accurate simulation of the multi-cavity modes in ACC39 we utilise the S3P suite of ACE3P. This does not have restrictions on the number of modes used per port. In all S3P calculations we consider all of the ports (and power couplers) in ACC39. No magnetic boundaries are applied at the ports. A selection of transmission results across ACC39 calculated using S3P are displayed in Fig 3a-3c.

In Fig 3a we clearly observe loss in power as the modes propagate down the module, particularly in bands that do not propagate across cavities such as: the first monopole band, the first beampipe band, the region just after the second dipole band and before the first quadrupole band. When we consider transmission across each subsequent cavity individually in the string (refer to Fig 3b) we observe that no two cavities have directly comparable results.

This occurs due to the modes being strongly coupled to successive cavities. In effect, the impedance loading of a particular cavity varies depending on the location of the cavity. This will of course make experimental determination of a particular eigenmode a non-trivial process. It is perhaps worth emphasing that these results are not an artefact of the numerical calculations, for if we consider the propagation across the beampipes in Fig 3c we obtain the expected spectrums (i.e. *c1p2,c2p1* should be identical to *c3p2,c4p1* since these geometries are the same [4])

## COMPARISON TO EXPERIMENT

There is reasonable agreement between the calculations made using GSM and S3P for an idealised ACC39 structure. However when compared to the experimental data (in which there was no beam present) we observe large discrepancies. This is illustrated in Fig 4.

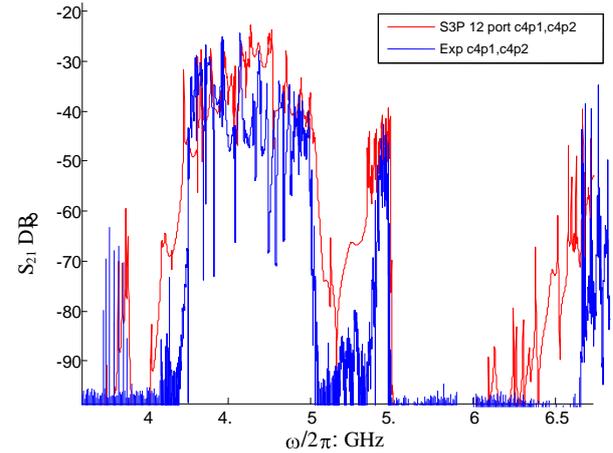

Figure 4: Direct comparison of simulated $S_{21}$ (S3P red) and the experimental data (DESY blue) across cavity C4 up to 7GHz (*c4p1,c4p2*).

If consider a single region such as the 5[th] dipole in which we know consists of predominantly localized modes, the discrepancy is not as large.

We believe that these discrepancies are due to the sensitivity of ACC39 to machining and alignment errors. This will be discussed in the next section.

## SENSITIVITY OF ACC39 TO ERRORS

A study on the sensitivity of the first monopole band to fabrication errors was conducted using a circuit model and a mode matching approach. The results of this investigation are detailed here [6]. This initial work considered only a single isolated cavity without couplers, and only investigated the monopole bands, in particular the fundamental band. It was found that the fundamental band of ACC39 was quite sensitive to fabrication errors.

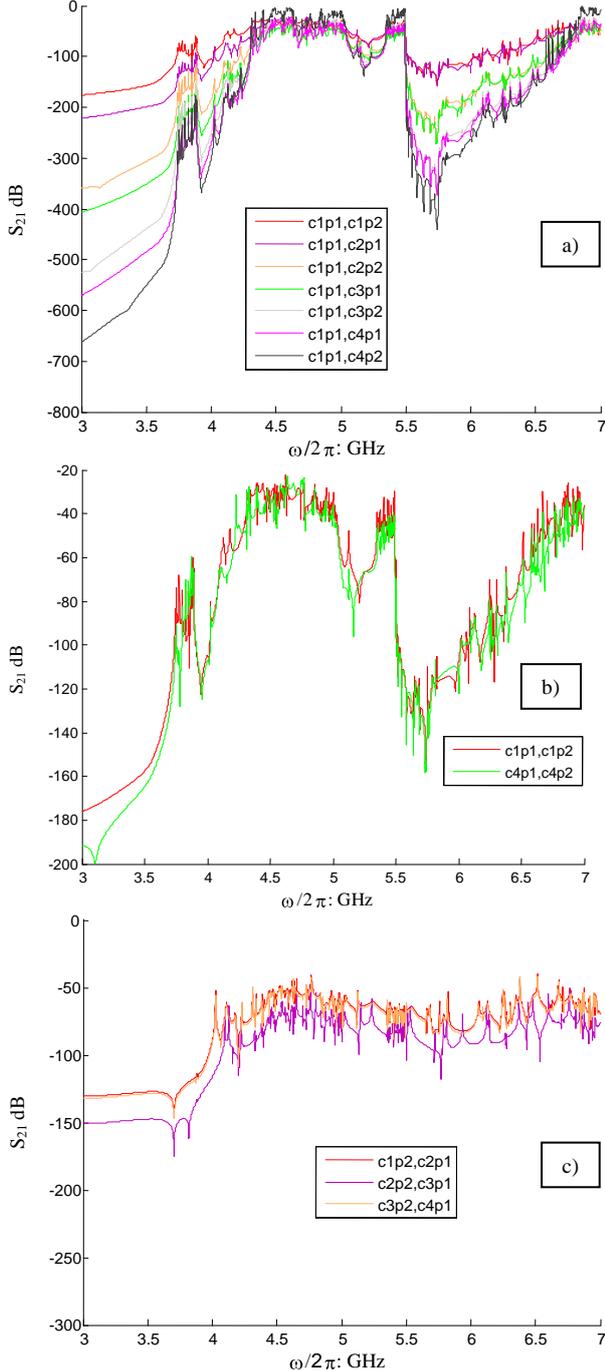

Figure 3: Comparison of $S_{21}$ transmission data across ACC39 calculated by S3P across various ports in ACC39.

It is computationally expensive to add random errors to ACC39 using S3P. Instead we utilise the GSM as it has been established that this technique is both rapid and reasonably accurate. Here we simulate random errors in the ACC39 structure by adding small pipe sections to the existing GSM unit cell. These pipe sections of random length are calculated analytically [7]. These sections serve the purpose of shifting the concatenated scattering matrices in a similar manner as to what would occur if geometric errors were included in the GSM unit cell calculations.

The RMS transmission ($S_{21}$) of 30 separate calculations with random seeds is displayed in Fig 5, in these calculations the length of the random pipe sections was varied with an rms of 0.35mm.

Sensitivity to these errors is evident. This underlines the results of Fig 4 in which a similar discrepancy is observed.

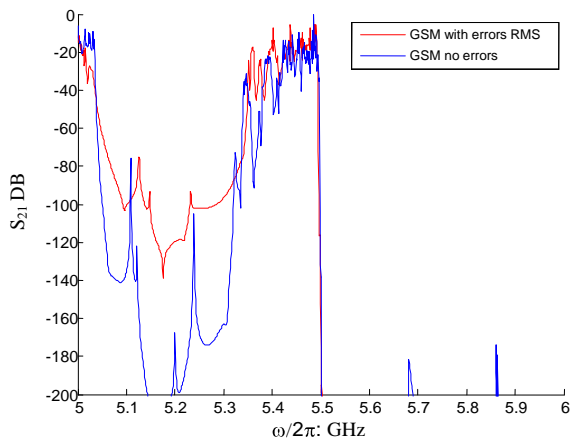

Figure 5: $S_{21}$ calculated by GSM using 20 modes per port across the entire module of ACC39 (*c1p1,c4p2*). Note here in this calculation all other ports are bounded by magnetic boundary conditions. Here we compare the idealised GSM calculation (blue) to that of the RMS of 30 GSM calculations (red) in which random errors with an RMS of 0.35mm have been added throughout ACC39 module.

The beampipe modes are of course particularly sensitive to errors as the bellows are known to be flexible and shift laterally by several mm. In contrast, the modes in the fifth dipole passband (9GHz to 9.1GHz) are trapped within the cavity and exhibit less sensitivity to machining and alignment errors.

## FINAL REMARKS

Simulations made with GSM and S3P have been shown to be comparable. However we note that in order for GSM to be accurate at least one propagating mode must be present. Furthermore, below the beampipe cutoff of ~4.39GHz, an accurate calculation using GSM requires a large number of evanescent modes

ACC39 is sensitive to fabrication and misalignment errors, particularly the second dipole and the first beampipe band. These misalignment errors shift the idealised multicavity resonances. Due to these largely unpredictable frequency shifts in the multicavity HOM's in ACC39 it becomes difficult to ascertain the exact nature of any given HOM. Despite this, the 5th dipole band remains a strong candidate for a HOM BPM system within ACC39 due to the localisation of the modes within this band and the diminished sensitivity of this region to fabrication and misalignment errors.

## ACKNOWLEDGEMENTS

This research has received funding from the European Commission under the FP7 Research Infrastructures grant no. 227579. We are pleased to acknowledge E. Vogel [8] for contributing important geometrical details on the cavities and coupler within ACC39. We would also like to acknowledge L. Xiao [5] for assistance with regards to the ACE3P computational suite.